%% file: mult_average_v1.tex
\documentclass{article}

\usepackage{epsfig}
\usepackage{graphicx}
\usepackage{array}
\usepackage{color}

\begin{document}
\pagestyle{empty}

\begin{flushright}
hep-ex/02 \\*[8mm]
\today
\end{flushright}
\vspace{1.8cm}

\centerline{
\parbox{12cm}{\bf
\begin{center}
\Large{Empirical parameterization of the high energy behaviour
       of average charged particle multiplicities in $e^+e^-$, 
       $ep$\/ and $pp$\/ collisions
      }
\end{center}
}}
\vspace*{1.7cm}

\begin{center}
\large{
M.Zavertyaev \\
Max-Planck Institut f\"ur Kernphysik, D-69117 Heidelberg, Germany \\
P.N.Lebedev Physical Institute,117924 Moscow B-333, Russia}
\end{center}
\vspace*{1.7cm}
\hrule
\vspace*{0.5cm}

\noindent{\bf Abstract} \\
          
A simple fit of existing data for center-of-mass energies above the
$b$-threshold shows that the average charged particle multiplicities 
in $p(\overline{p})$\/ and $pp$\/ interactions can be parameterized
by a simple power law $\langle n_{ch} \rangle= N_0 \cdot (s/s_{0})^{\gamma}$\/
with $\gamma$ close to $1/5$. Choosing $s_0=m_p^2$\/ one finds $N_0$=2.32. 
The average charged particle multiplicities in $e^+e^-$\/ and $e^{\pm}p$ 
interactions follows the same power law with an offset of $\pm 1$\/
unit in $N_0$. The observed trend leads to the conclusion that in LHC 
experiments at energy of 7\,TeV one should expect the average charge 
multiplicity in the region of $\langle n_{ch} \rangle$ close to 80. 

\vspace*{0.5cm}
\hrule

\vfill

\clearpage
\large

Many experiments published their results on average charge multiplicities studies
at different energies in $e^+e^-$, $e^{\pm}p$  and $p(\overline{p})$\/ and $pp$\/ 
interactions. A compilation of these data may be found in a data files prepared
by Particle Data Group (\cite{PDG} and references therein). 


In Fig.\,\ref{fig:PDG} the data taken from \cite{PDG} are presented 
in double logarithmic scale. For $\sqrt{s}>11$\,GeV, just above the mass
of the Upsilon resonances, the  $p(\overline{p})$\/ and $pp$\/ data are 
fitted by  $\langle n_{ch} \rangle = N_0 \cdot (s/m_p^2)^{1/5}$\/ with $m_p$\/
the proton mass and one free parameter $N_0$. Such a simple power law may 
find its explanation if 
the process of multiparticle production has some sort of fractal nature 
in it. It is intriguing that the power $1/5$\/ which provides a very 
good fit to the high energy data appears to be equal to $1/n_f$, the 
number of active flavours in the process. Furthermore, the best fit 
value $N_0=2.32$\/ is very close to $\ln(m_{\Upsilon}/m_p)$.

Another curious observation is related to the $e^+e^-$\/ and the 
$e^{\pm}p$\/ data points. The green and the red lines are obtained from 
the fit to the $p(\overline{p})$\/ and $pp$\/ data by changing $N_0$\/
by $\pm 1$\/ unit. Surprising is the precision with which the data points 
from $e^+e^-$\/ and $e^{\pm}p$ interactions follow the same power law.
One also sees that the $e^{\pm}p$\/ measurements, which pertain to the 
current fragmentation region in the Breit-frame, are about half of the 
$p(\overline{p})$\/ and $pp$\/ multiplicities.

The observed trend leads to the conclusion that in LHC experiments at energy
of $\sqrt{s} $ = 7\,TeV one should expect the average charge multiplicity in 
the region of $\langle n_{ch} \rangle$ close to 80 - the extrapolation for about 
one order of magnitude of energy. In Fig.\,\ref{fig:PDG} the expectation region
marked by a big magenta dot. 

It would be very interesting to see the results on average charge multiplicity 
in the future experiments on $e^+e^-$ and $e^{\pm}p$ new generation
accelerators.

\section*{Acknowledgments}
\label{ackn}

The author acknowledges very useful discussions with M.Schmelling.

\clearpage

\input{fig1.tex}

\end{document}

%% file: fig1.tex
\begin{figure*} 
\addtolength{\abovecaptionskip}{10pt}
\centering
\includegraphics[width=15cm]{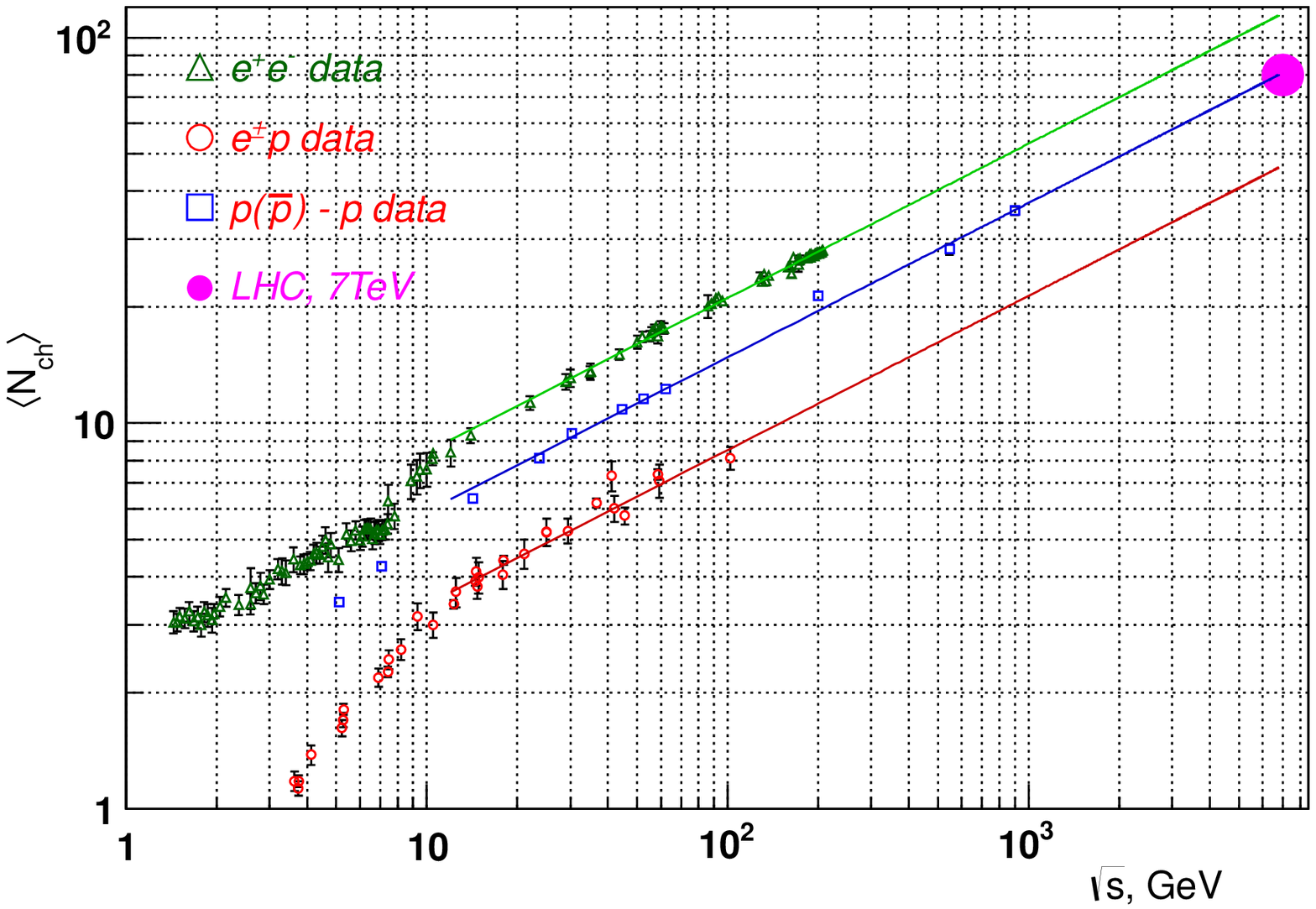}
\caption{ 
         The average charge multiplicities in $e^+e^-$, $e^{\pm}p$ and 
         $p(\overline{p})$\/ and $pp$\/  interactions. The data points are 
         taken from  \protect\cite{PDG}. The blue line is the result of
         the fit of data in $p(\overline{p})$\/ and $pp$\/  interaction
	 with the power fixed to 1/5, the 
	 green and read lines are the offset of fit by $\pm 1$ unit.
	 The magenta dot indicates the expected  average charge 
	 multiplicities at 7\,TeV.
	 }
\label{fig:PDG}
\end{figure*} 

%% file: mult_average_v1.bbl
\newpage
\begin{thebibliography}{}

\bibitem{PDG}
Nakamura, K. {\it et al., Review of particle physics, J.Phys., 2010.}

\end{thebibliography}
